\documentclass[traditabstract,letter]{aa}  
\usepackage{amsmath,amssymb}
\usepackage{natbib}
\usepackage{txfonts}
\usepackage{graphicx}
\bibpunct{(}{)}{;}{a}{}{,} 

\def\m2s2{\hbox{\,m$^{2}$\,s$^{-2}$}} 
\def\Msun{\hbox{$M_{\odot}$}}             
\def\Rsun{\hbox{$R_{\odot}$}}

\def\chisq{\mbox{$\chi^2$}}

\def\target{HD~144548}

\begin{document}

\title{HD144548: A young triply eclipsing system in the Upper Scorpius OB association \thanks{Partially based on observations made with the Italian Telescopio Nazionale Galileo (TNG) operated by the Fundaci\'on Galileo Galilei of the INAF, the Nordic Optical Telescope, operated by the Nordic Optical Telescope Scientific Association, and the William Herschel Telescope (program DDT58 - PI Lodieu) operated by the Isaac Newton Group on the island of La Palma at the Spanish Observatorio Roque de los Muchachos of the IAC.}}

\author{Alonso, R. \inst{1,2}
\and Deeg, H.J. \inst{1,2}
\and Hoyer, S. \inst{1,2}
\and Lodieu, N. \inst{1,2}
\and Palle, E. \inst{1,2}
\and Sanchis-Ojeda, R. \inst{3,4}
}


\institute{
Instituto de Astrof\'\i sica de Canarias, E-38205 La Laguna, Tenerife, Spain
\and Dpto. de Astrof\'isica, Universidad de La Laguna, 38206 La Laguna, Tenerife, Spain\label{La Laguna}
\and Department of Astronomy, University of California, Berkeley, CA 94720
\and NASA Sagan Fellow
}

\date{Received 3 August 2015 / Accepted 12 October 2015}
\abstract
{The star \target\ (=HIP~78977; TYP~6212-1273-1) has been known as a detached eclipsing binary and a bona-fide member of the Upper Scorpius OB association. Continuous photometry from the K2 mission on Campaign Two has revealed the presence of additional eclipses due to the presence of a third star in the system. These are explained by a system composed of the two previously known members of the eclipsing system ($Ba$ and $Bb$) with a period of 1.63~d, orbiting around an F7-F8V star with a period of 33.945$\pm$0.002~d in an eccentric orbit ($e_A$ = 0.2652$\pm$0.0003). The timing of the eclipses of $Ba$ and $Bb$ reveals the same 33.9~d periodicity, which we interpret as the combination of a light time effect combined with dynamical perturbations on the close system. Here we combine radial velocities and analytical approximations for the timing of the eclipses to derive masses and radii for the three components of the system. We obtain a mass of 1.44$\pm$0.04 \Msun\ and radius of 2.41$\pm$0.03 \Rsun\ for the $A$ component, and almost identical masses and radii of about 0.96 \Msun\ and 1.33 \Rsun\ for each of the two components of the close binary. \target\ is the first triply eclipsing system for which radial velocities of all components could be measured.}


\titlerunning{HD144548}
\authorrunning{All et al.}

\maketitle

\section{Introduction}
\label{sec:intro}
Triply eclipsing stars are very rare systems, in which all components of a triple hierarchical stellar system are mutually eclipsing. When the three stars are in a compact configuration, they are excellent laboratories to check dynamical effects that have shorter timescales than in stellar binaries, and from which information on the stellar interiors can be inferred (e.g. \citealt{Claret:1993aa}). If the system is observed photometrically during a long period of time, it is possible to measure the physical parameters of the three stars with remarkable accuracy. KOI-126 was the first triply eclipsing system found \citep{Carter:2011ab}, and it allowed to determine the masses and radii for two of the components down to 3\% and 0.5\% fractional uncertainty, respectively, with a combination of a photo-dynamical model and radial velocity observations. Two other triply eclipsing systems have been found thanks to the Kepler mission (HD~181068, \citealt{Derekas:2011aa}; KIC002856960, \citealt{Armstrong:2012aa}, \citealt{Lee:2013aa}), while 26 other triple systems (the majority not showing eclipses of the outer star) have been detected thanks to the study of the Eclipse Time Variations (\citealt{Rappaport:2013ac,Borkovits:2015aa}). 

Upper Scorpius (USco) is part of the nearest OB association (145 pc). The central region of USco is free of extinction and star formation has already ended \citep{Walter:1994aa}. The age of USco remains under debate. \cite{Preibisch:1999aa} originally estimated an age of 5-6~Myr with a small dispersion based on the location of pre-main sequence members in the HR diagram. Recently, \cite{Pecaut:2012aa} revised the age of the isochronal ages of intermediate-mass members to 11$\pm$2~Myr, in contrast to the 4$\pm$1~Myr age derived for low-mass members (\citealt{Slesnick:2008aa,Lodieu:2011aa,Herczeg:2015aa}).
For \target\ most models suggest an age of 8-11 Myr, its kinematic distance is 133$\pm$12~pc, and it is classified as a F8V star in USco with an effective temperature of 6138K and an Av of 0.38$\pm$0.08 mag (Table 4 of \citealt{Pecaut:2012aa}).
\target\, is an astrometric member of USco based on Hipparcos data (\citealt{de-Zeeuw:1999aa,Pecaut:2012aa}), with reported detection of Lithium and 24~$\mu$m excess due to a debris disk \citep{Chen:2011aa}, and it is a known X-ray source from ROSAT data \citep{Haakonsen:2009aa}. 
Using ASAS photometry of ROSAT sources, \target\ had been identified as a detached eclipsing binary system by \cite{Kiraga:2012aa}, with a reported period of 0.81292~d, although they note a possible period twice as long. In this work, we identify a period of 1.63~d for the orbit of a close binary system, further denoted binary $1$ with components $Ba$ and $Bb$, which is itself orbiting around a longer period (33.9~d) eclipsing component (binary $2$, component $A$), forming a compact hierarchical triply eclipsing system.

\section{Observations and data analysis}
\label{sec:obs}

\begin{figure*}
\centering
\includegraphics[width=18cm]{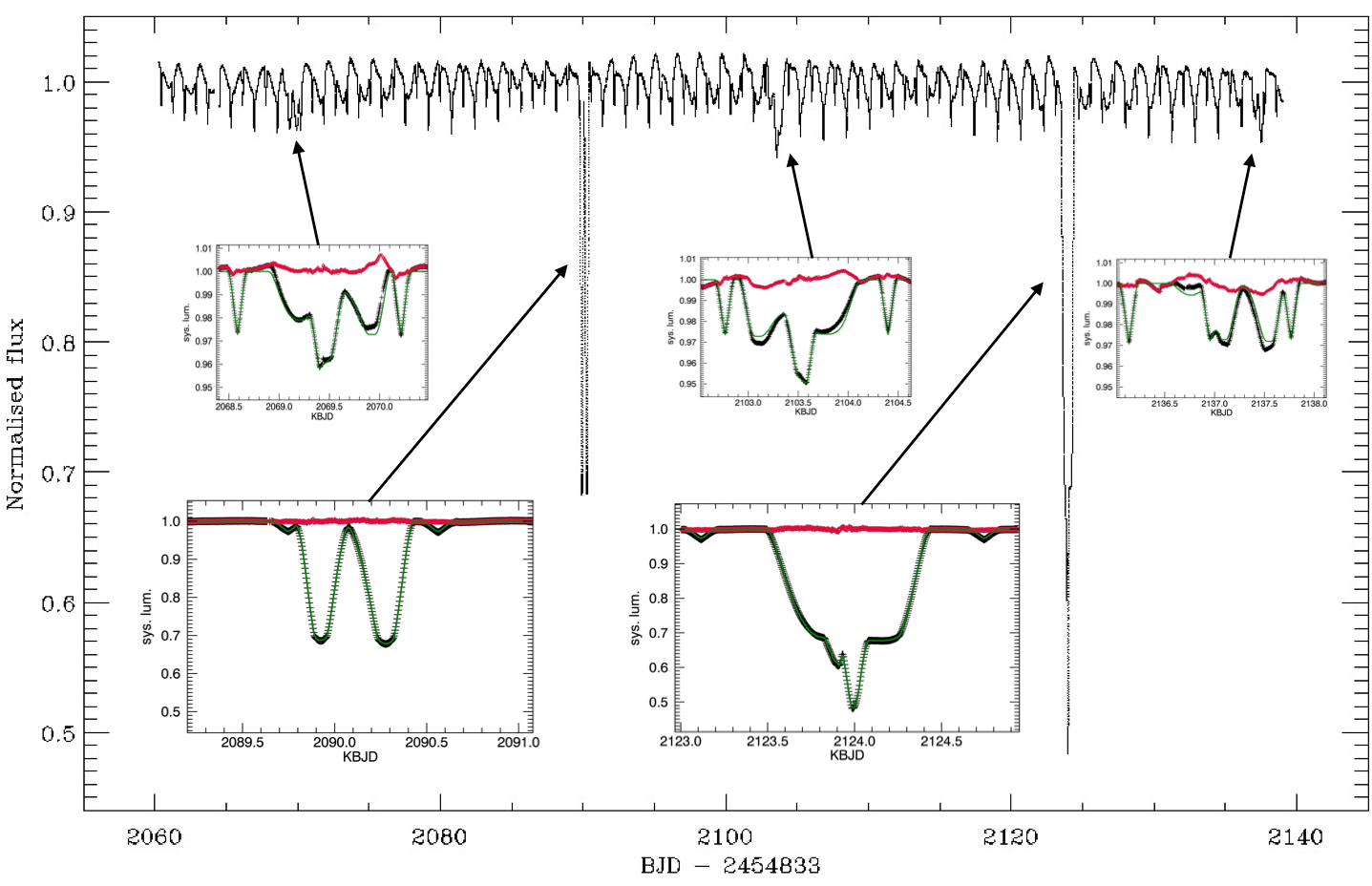}
\caption{The extracted K2 light curve of \target, showing periodic eclipses of a hierarchical triple eclipsing system, and clear modulation that we attribute to the effect of rotation and spots. The insets show primary and secondary eclipses of $A$ vs. $B_a$ and $B_b$ components, with the best fit model (green dotted lines) and residuals of the fit (red crosses, shifted vertically to the off-eclipse flux).} 
\label{fig:fig_lc}
\end{figure*}

The K2 \citep{Howell:2014aa} Campaign Two data\footnote{Proposed for observations in long cadence mode ($\sim$30~min cadence) by Guest Observer programs 2020, 2021, 2023, 2034, 2049, 2052, 2086, 2092, and in short cadence mode ($\sim$1~min cadence) by program 2023.} for the star \target\ (EPIC-204506777) spans 78.78~d, starting on Aug 23, 2014. The star is bright (V=8.75, Kepmag = 8.75), and appears slightly saturated in the target pixel files.

We extracted the photometry from the short cadence data, which consists on 115680 windows with a size of 16x29 pixels approximately centered on the target star. For each time stamp, we determined the centroid by performing gaussian fits to the mean flux level at each spatial direction, to avoid two dimensional fits that may get confused by the changing vertical trails arising from the saturation of the images. A circular aperture with a radius of 7 pixels was used to obtain the flux from the target. The columns close to the determined X-center of the target were summed in their whole length, to include the flux that was dispersed due to saturation. The background level was estimated from an 8.4 pixel-wide annulus starting 9.1 pixels away from the measured centroid. Several outliers due to cosmic rays impacts were located and interpolated through a 5-sigma clipping moving median window. Less than 0.8\% of the data points were corrected this way. The final light curve is plotted in Figure~\ref{fig:fig_lc}, where two deep eclipses of the close binary passing before component A are dominant in the curve, about 34 days apart. Also detected are three occultations of the binary star behind the star $A$. The mutual eclipses of the $Ba$-$Bb$ system are seen throughout the whole light curve, with a period of 1.6278~d.

The light curve exhibits a modulation of few percent between eclipses of the close components. The autocorrelation function of this modulation shows a maximum at a period close to, but not exactly, the orbital period of the close binary $1$.  A division of the curve used to compute the autocorrelation by a 20 points median-smoothed version of it is used to estimated the dispersion of the residuals as 210~ppm.

\subsection{Observed minus Calculated (O-C) diagram}
\label{sec:sec_omc}

The highly significant individual eclipses of close binary $1$ allow for a timing of their center positions. The timing of the eclipses departs from a strict periodicity due to a combination of the Light Travel Time Effect (LTTE) and several dynamical effects that change the intrinsic period of binary 1. For a complete description of the dynamical effects, we refer the reader to the work of \cite{Borkovits:2015aa}, hereafter B15. To describe the O-C diagram, we implemented the analytical equations of that work, including the following terms: i) the LTTE, ii) dynamical effects with a timescale of the outer binary period up to second order ($\Delta_1$, Eq. 5 of B15, and $\Delta_2$, Appendix A of B15). We added a linear trend on the argument of periastron $\dot{\omega}_2$, to account for a possible apsidal rotation.

\begin{figure}
\centering
\includegraphics[width=\columnwidth]{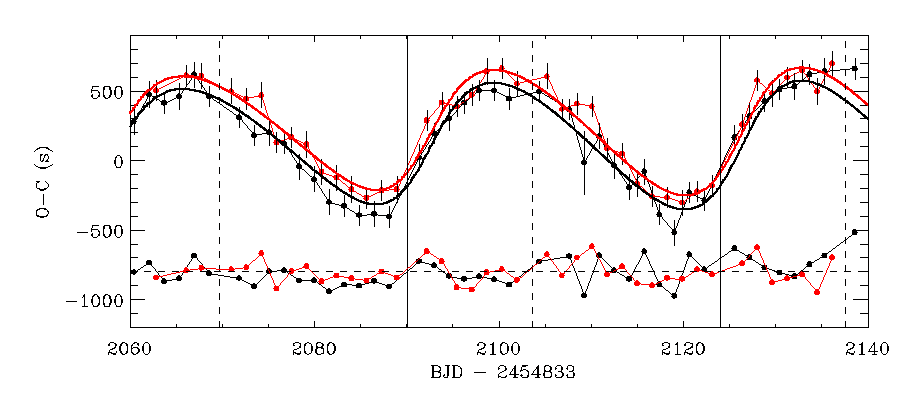}
\caption{Observed minus calculated diagram, and best-fit model. Red: primary eclipses, black: secondary eclipses. The vertical lines mark the positions of the eclipses and occultations of the outer binary 2.} 
\label{fig:fig_omc}
\end{figure}

The preparation of the light curve and the extraction of the O-C diagram is described in further detail in Appendix~\ref{sec:appenA}. The result is plotted in Figure~\ref{fig:fig_omc}. We performed an initial fit of the O-C times using the {\sc amoeba} minimization algorithm \citep{Press:1992aa}, that included 18 parameters to the fit. As the number of data points is comparable to the model parameters, the fit is unconstrained, and at this stage it was used only to prepare the data for the final modelling described in Sect.~\ref{sec:ana} 
  
\subsection{Radial velocities}
\label{sec:rv}

While a fit to the O-C in the previous section is in principle the equivalent of observing a double-lined spectroscopic binary, an increased precision on the orbital parameters of the system is obtained with radial velocity measurements \citep{Carter:2011ab}. The system \target\ was observed at the 2.5~m NOT with the FIES spectrograph \citep{Telting:2014aa} on the night of May 7th, 2015. The exposure time was 1800s.  Four additional measurements were obtained with the HARPS-N spectrograph at the 3.6~m TNG on four consecutive nights starting on May 21st, 2015, using an exposure time of 900~s and reaching a S/N in the red part of around 130 per spectrum. An additional spectrum was taken with the ISIS spectrograph at the William Herschel Telescope on June 8th, 2015. For the FIES and ISIS data points, the spectra were reduced and extracted with regular tools (the online data reduction {\sc FIESTool}\footnote{http://www.not.iac.es/instruments/fies/fiestool/}, and {\sc IRAF}\footnote{http://iraf.noao.edu/}). The radial velocities were obtained with IRAF {\sc fxcor}, and standard radial velocity stars observed on the same night were used to obtain the absolute velocities. 

\begin{figure*}
\centering
\includegraphics[width=\textwidth]{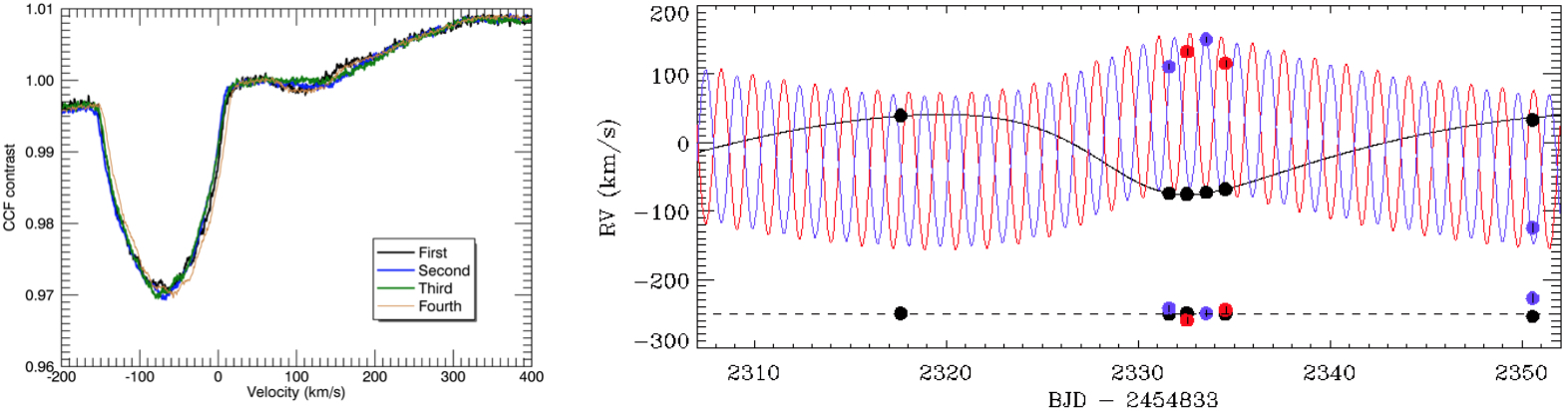}
\caption{Left: the HARPS-N CCFs obtained on four consecutive nights, showing a wide main component at about -80~km/s, and more subtle CCFs around 100~km/s. Right: the best fit orbital solution for the three components of the system (black: A, blue: Ba, red: Bb), and the residuals of the fit shifted for clarity (points near dashed line). These are $\sim$1.6~km/s for the $A$ component, and $\sim$11~km/s for each of the components of binary 1.} 
\label{fig:fig_rvs}
\end{figure*}

For the HARPS-N spectra, we obtained the cross-correlation functions (CCFs) over a 400 km/s velocity region using the online YABI tool (\citealt{Borsa:2015aa,Hunter:2012aa}), using a G2V mask. The CCF is dominated by a broad ($\sim$160~km/s) peak, and shows signs of an additional CCF component at velocities around +100~km/s (Figure~\ref{fig:fig_rvs}). We attribute these secondary peaks to one of the components of the close binary system. Using the ephemeris of the three components, and making first reasonable guesses on the masses of each system component, we estimated the position of the CCF of the second component of the close binary system to lie blended with the right wing of the CCF of the A component of the system. According to this, the first obtained HARPS-N spectrum is the least contaminated by this component, and we used its velocity as a reference point. For the other HARPS-N spectra, we estimated the radial velocity shifts via a cross-correlation of the CCFs using only the left wing of the reference CCF. For the secondary CCFs visible around 100~km/s, we estimated their central position by a baseline normalization and a fit to a gaussian, using a K5V correlation mask. The estimation of the center of the secondary CCF using two different functions (gaussian and a trapezoid) was used to evaluate the error bars, which are summarized in Table~\ref{tab:tab_rvs}, and the best orbital fit to the three components of the system is plotted in Figure~\ref{fig:fig_rvs}. To perform this fit, we limited the rate of apsidal motion $\dot{\omega}_2$ of binary 2 to the values obtained from the light curve fitting described in the following section, as this quantity is better constrained by the ephemeris of the primary and secondary eclipses of the $A$ component.

\subsection{Light curve modelling of the eclipses}
\label{sec:ana}

Fits of the stellar photometric eclipses were obtained with a revised version of the UFIT fitter program using the {\sc amoeba} algorithm against models generated with UTM (Universal Transit modeller, both \citealt{Deeg:2014aa}). UTM permits the generation of light curves from any number of mutually occulting objects (stars, planets, moons, planetary rings) on arbitrary elliptical orbits in hierarchical systems. The bodies' orbital positions are based on the usual elliptical orbital parameters: period $P$, semimajor axis $a$, eccentricity $e$, argument of periapsis $\omega$, inclination $i$, longitude of ascending node (or position angle) $\Omega$, time relative to a reference epoch, and apsidal motions $\dot{\omega}$. 

In a first fitting-pass, we used the average phased curves of binary $1$ shown in Fig.~\ref{fig:fig_pri_sec_model} to derive this binary's relative radii $R_{Ba}/a_1$ and $R_{Bb}/a_1$, radius ratio $k_1 = R_{Bb}/R_{Ba}$, Luminosity ratio $q_{L1}=L_{Bb}/L_{Ba}$, inclination $i_1$ and the fraction of  its luminosity relative to the total luminosity, with all other light assumed to be from the component $A$. For the limb-darkening, we coupled the coefficients of the two very similar components and fitted only a linear coefficient, since neither individual coefficients nor higher-order limb darkening resulted in significantly better fits. The residuals of these fits to binary $1$, of 0.0031 in normalised flux units, are dominated by asymmetries in their eclipse events which are likely due to inhomogeneous surfaces of these stars.

In the second pass, we concentrated on the eclipses involving component $A$. For these fits, a light curve was used in which each time-stamp was corrected for the O-C time resulting from the fit to O-C times described in Sect.~\ref{sec:sec_omc}. Thus, the shallow eclipses of the binary $1$ appear with strict periodicity. The light curve was also median-filtered and rebinned by a factor of 5. Furthermore, we trimmed the light curve to include only sections around the $A$-component eclipses, extended to include one binary-$1$ eclipse on either side. Parameters derived from the fit to binary $1$ were initially kept fixed, though we permitted small variations in the final fits. Linear limb darkening parameters were used. As expected, the eccentricity $e_2$ and the argument of periapsis $\omega_2$ were strongly correlated. We proceeded therefore with fits in which we gave as initial conditions an estimate of the phase $\phi_{s,2}$ of secondary eclipses on component $A$ and for a given value \textbf{$\omega_2$} (that was kept fixed), calculated an initial eccentricity $e_2$. Fits were then repeated for sets of fixed values of  $\omega_2$. 

\subsection{Combined fit}

We used the obtained orbital inclinations, periods, phase of secondary eclipse of component $A$, and apsidal motion to perform a combined fit of the radial velocity and O-C curves, using these as priors. In the fits, we also changed slightly the initial values of the remaining parameters, and 4000 {\sc amoeba} chains were computed, which converged after $\sim$2000 iterations per chain. The distributions of all the iterations with a \chisq\ lower than the final \chisq\ of the worst converging chain were used to estimate the final error bars.  

The best light curve fit, for which the values in Table~\ref{table:param} are given, has an overall residual $rms$ of 0.0021 in normalized flux units within the main eclipses involving component A (insets of Figure~\ref{fig:fig_lc}), and 0.0017 for the rest of the light curve. We note that these residuals are an order of magnitude smaller than the $\sim$3\% peak-to-peak variability of the raw light curve (Figure~\ref{fig:fig_lc}). The major origin of the residuals is probably an imperfect removal of that variability, which is likely caused by surface inhomogeneities of the stellar surfaces on any of the components. During eclipse events, these inhomogeneities may cause further brightness variations on time-scales that are different than the off-eclipse variations, and which cannot be corrected for. Another source of deviation comes from limits in the model-generation, which is based on Keplerian orbits plus apsidal motion. The dominant source of the observed O-C residuals are dynamical effects on shorter time-scales, which ultimately arise from deviations of the bodies' orbits from Keplerian ones. Such deviations against the modelled positions of the bodies may of course affect the modelled eclipse light curves. A full dynamical 3-body simulation of the system, as was done with KOI-126 \citep{Carter:2011ab}, or a conversion of the analytical terms describing the O-C deviations in the framework of B15 into geometric deviations might remedy this. This is however beyond the scope of this work and given the rather large flux variations of the system not arising from eclipses, it might not lead to a relevant improvement in the knowledge of the system's parameters.

\section{Conclusions}
\label{sec:concl}

We report the discovery of a young (8-11~Myr) triply eclipsing system member of the nearest OB association to the Sun. The physical parameters obtained for the each of the components of the system are good calibrators for stellar models at young ages. The obtained radii of the stars are about 65\% and 40\% larger than main sequence stars for the component $A$ and each of the components $Ba-Bb$, respectively, as is expected due to the young age of the system. We note that \target\ is the youngest and one of the most compact triple systems known, the other being KOI-126 (outer period 33.92~d) and $\lambda$-Tau (outer period 33.03~d, \citealt{Tokovinin:2008aa}). This accumulation near P$_2\sim$33~d might hint about dynamical limits in the origin of these compact triple systems. From the difference between inclinations ($i_1 - i_2$)  and from $\Delta P.A.$, we conclude that the orbital planes are coplanar within 2~$\deg$.

Given the brightness of the system, we expect that further observations of eclipses of component $A$ will be a rewarding topic for observatories even with modest equipment. Additional radial velocities at $\sim$km/s precision might also help to calibrate some possible remaining systematics, such as the effect of the stellar activity or the extraction of the RVs for the close binary 1, and refine the parameters described in this work, especially the dynamical effects. Finally, a determination of the spin-orbit alignment of the component $A$ vs. component $Ba-Bb$ and between both components of the close system from observations of the Rossiter/McLaughlin effect would provide constraints to tidal forces on young stars. 

\begin{acknowledgements}
We thank the anonymous referee for useful comments that helped to improve the paper. 
RA and NL were funded by the Ram\'on y Cajal fellowship (RYC-2010-06519 and 08-303-01-02, respectively). HD and SH acknowledge support by grant AYA2012-39346-C02-02 and under the 2011 Severo Ochoa Program SEV-2011-0187, both from the Spanish Secretary of State for R\&D\&i (MINECO). RA, EP acknowledge funding from MINECO grants ESP2013-48391-C4-2-R and ESP2014-57495-C2-1-R. RSO acknowledges NASA/JPL funding through the Sagan Fellowship Program executed by the NASA Exoplanet Science Institute. We thank the WHT staff and its service program, the TNG staff, the NOT staff and Jorge P. Arranz for help during the observations. 
\end{acknowledgements}

\onltab{
\begin{table}[h]
\caption{Radial velocities}            
\centering        
\begin{tabular}{lcrrr}       
\hline\hline   
BJD -& Source & RV$_A$  & RV$_{Ba}$ & RV$_{Bb}$ \\ 
2450000 & &[km/s] &[km/s] & [km/s]\\
\hline
7150.63672 & FIES & 39.1$\pm$0.4 & -- & -- \\
7164.62688 & HARPS-N & -74.2$\pm$0.2 & 110.7$\pm$4.0 & --\\
7165.55411 & HARPS-N & -75.7$\pm$0.2 & -- &132.7$\pm$4.0\\
7166.56063 & HARPS-N & -73.0$\pm$0.2 & 150.3$\pm$4.0 & --\\
7167.55627 & HARPS-N & -68.5$\pm$0.2 & -- & 115.7$\pm$4.0 \\
7183.56465 & ISIS &  32.7$\pm$3.4 & -125$\pm$6.0 & -- \\
\hline
\end{tabular}
\label{tab:tab_rvs}    
\end{table}
}
 
\onltab{
\begin{table*}[h]
\caption{ Parameters of the triple system.}            
\centering        
\begin{tabular}{lr}       
\hline\hline   
\\  
\multicolumn{2}{l}{Light-curve fit of binary $1$ mutual eclipses} \\         
\hline \\
period, $P_1$ (d) & 1.6278 $\pm$ 0.0001\\
epoch$^1,$ $T_{1}$\ (BKJD)$^2$ & 2061.2610 $\pm$ 0.0012 \\    
inclination, $i_1$ (deg) & 88.9 $\pm $ 0.5\\
eccentricity, $e_1$ & $\lesssim$ 0.0015\\
relative star radius, $R_{Ba}/a_1$ & 0.1834 $\pm$ 0.0012\\
radius ratio, $k_1 = R_{Bb}/R_{Ba}$ & 1.008 $\pm$ 0.005\\
comp. $Ba, Bb$ linear limb-dark. coeff., $u_a$ & 0.78 $\pm$ 0.07\\
luminosity ratio, $L_{Bb}/L_{Ba}$ & 0.97 $\pm$ 0.05 \\
\\
\multicolumn{2}{l}{Light curve fit of binary $2$ (eclipses with component $A$)} \\   
\hline \\
period, $P_2$ (d)  & 33.945 $\pm$ 0.002\\
epoch$^3$, $T_2$\ (BKJD) & 2090.0680 $\pm$ 0.0010\\  
inclination,  $i_2$ (deg) & 89.28 $\pm$ 0.05\\
relative position angle$^5$, $\Delta P.A.$ & -1.0 $\pm$ 0.9\\
secondary eclipse phase$^4$, $\phi_{s,2}$ & 0.3950 $\pm$ 0.0005\\ 
eccentricity, $e_2$ &  0.2652 $\pm$ 0.0003\\
arg. of periapsis$^6$, $\omega_2$(deg) & 127.63 $\pm$ 0.15\\
apsidial rotation velocity, $\dot{\omega}_2$ (deg d$^{-1}$) & 0.0235 $\pm$ 0.002\\
epoch of periapsis, $T_{\nu,2}$(BKJD) & 2092.31 $\pm$ 0.02\\
relative star radius, $R_{A}/a_2$ & 0.0364 $\pm$ 0.0005\\ 
comp. $A$ limb-dark. coeff., $u$ & 0.37 $\pm$ 0.10\\
luminosity ratio, $(L_{Ba}+L_{Bb})/L_A$ & 0.0593 $\pm$ 0.0012\\
mass ratio $q_{12}=(m_{Ba}+m_{Bb})/m_A$ &  1.34 $\pm$ 0.03\\
mass ratio $q_1=m_{Bb}/m_{Ba}$ & 0.995 $\pm$ 0.015\\
\\
\multicolumn{2}{l}{Parameters from O-C and RV fit} \\          
\hline 
\\
amplitude of light travel time effect, $A_{LTTE}$ (s) & 86.7 $\pm$ 0.1 \\
amplitude of first order dynamical effects with P$_2$ timescale, $A_{L1}$ (s) &  957 $\pm$ 8 \\
amplitude of second order dynamical effects with P$_2$ timescale, $A_{L2}$ (s) &  2.4 $\pm$ 0.9 \\
mutual orbital inclination $i_m$ (degrees)& 0.2 $\pm$ 0.5\\
RV semi-amplitude $K_A$ (km/s) & 58.52 $\pm$ 0.05\\
RV semi-amplitude $K_{Ba+Bb}$ (km/s) & 43.6 $\pm$ 0.6\\
RV semi-amplitude $K_{Ba}$ (km/s) & 110.2 $\pm$ 1.2\\
RV semi-amplitude $K_{Bb}$ (km/s) & 115.0 $\pm$ 0.6\\
RV offset, gamma (km/s)& -6.97 $\pm$ 0.10\\
mass ratio  $q_{12}=(m_{Ba}+m_{Bb})/m_A$ &  1.342 $\pm$ 0.019\\
mass ratio  $q_{1}=(m_{Bb}/m_{Ba})$ &  0.959 $\pm$ 0.015\\
\\
\multicolumn{2}{l}{Adopted parameters} \\
\hline
\\
semimajor axis, $a_A$ (A.U.) & 0.17619 $\pm$ 0.00012\\
semimajor axis, $a_{Ba+Bb}$ (A.U.) & 0.1313 $\pm$ 0.0018\\
semimajor axis, $a_{Ba}$ (A.U.) & 0.0165 $\pm$ 0.0002\\
semimajor axis, $a_{Bb}$ (A.U.) & 0.01721$\pm$ 0.00009\\
semimajor axis, $a_{1}$ (A.U.) & 0.03371 $\pm$ 0.00011\\
semimajor axis, $a_{2}$ (A.U.) & 0.3075 $\pm$ 0.0013\\
mass component $A$, $M_A$ (\Msun) & 1.44 $\pm$ 0.04\\
mass component $B_a$, $M_{Ba}$ (\Msun) & 0.984 $\pm$ 0.007\\
mass component $B_b$, $M_{Bb}$ (\Msun) & 0.944 $\pm$ 0.017\\
equatorial radius component $A$, $R_A$ (\Rsun) & 2.41 $\pm$ 0.03\\
radius component $B_a$, $R_{Ba}$ (\Rsun) & 1.319 $\pm$ 0.010\\
radius component $B_b$, $R_{Bb}$ (\Rsun) & 1.330 $\pm$ 0.010\\
\\
\hline\hline
\end{tabular}
\label{table:param}    
\tablefoot{$^1:$  Epoch of the primary (deeper) eclipse, when component $Ba$ is being occulted by $Bb$.\\
$^2:$ Barycentric Kepler Julian Date,  BKJD = BJD - 2\ 454\ 833. \\
$^3:$ Epoch when component A would become eclipsed by the barycenter of the compact binary $1$. \\
$^4:$ Orbital phase of component A when the barycenter of components $Ba,Bb$ eclipses behind it. \\
$^5:$ Position angle of orbital plane of binary 2 relative to orbital plane of binary 1.\\
$^6:$ This value changes with time due to $\dot{\omega}_2$ and is given for the moment of the epoch $T_{\nu,2}$. \\
}
\end{table*}
}

\bibliographystyle{aa}
\bibliography{references}

\Online

\begin{appendix}

\section{Obtaining the O-C diagram}
\label{sec:appenA}

\begin{figure}
\centering
\includegraphics[width=\columnwidth]{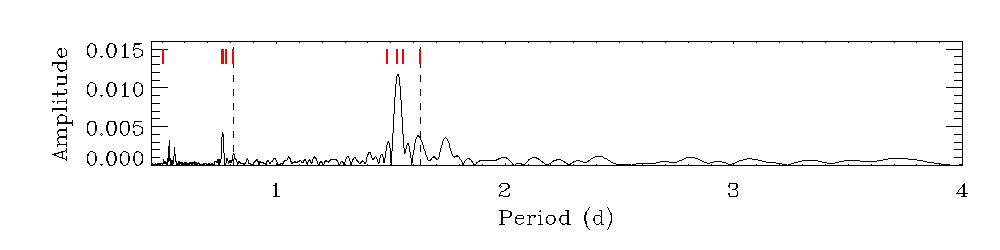}
\caption{Amplitude periodogram of the light curve shown with all the eclipses removed and interpolated. The vertical dashed lines mark the period of the close binary $1$ and its harmonics. The red thicks mark the position of the frequencies removed for further analysis.} 
\label{fig:fig_p04_spect}
\end{figure}

To prepare the light curve for the analysis, we first cleaned the most significant frequencies in the amplitude spectrum of the off-eclipse sections, using {\sc period04} \citep{Lenz:2005aa}. Twelve sinusoidal components were used, until amplitudes of the peaks of about 3 times the dispersion of the residuals were reached. Figure~\ref{fig:fig_p04_spect} shows the amplitude spectrum and the frequencies that were removed. As in the case of the autocorrelation, the spectrum shows frequencies that are close to, but not exactly, the one of binary $1$, and also shows peaks at the orbital frequency of the binary 1 and its harmonics. These might be caused by a combination of reflected light, ellipsoidal variability, and doppler beaming whose detailed analysis is out of the scope of this paper. 

\begin{figure}
\centering
\includegraphics[width=\columnwidth]{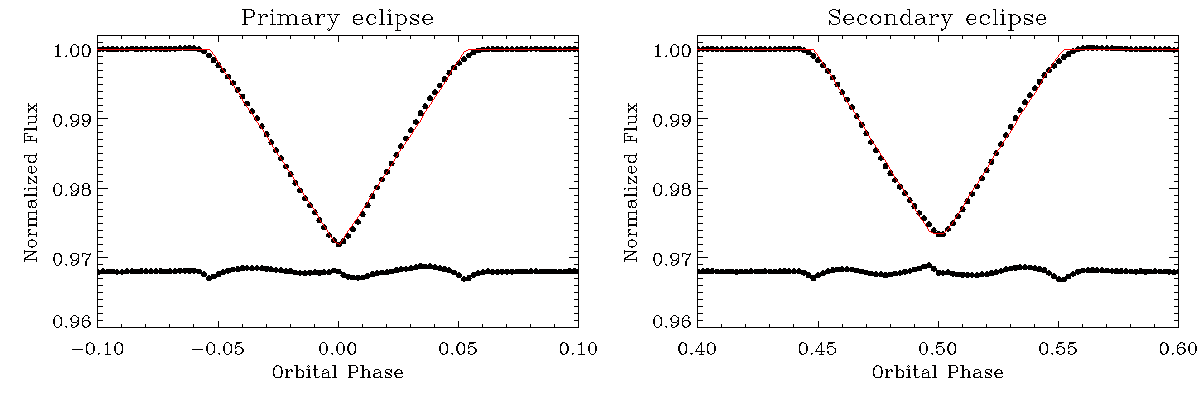}
\caption{Averaged curves of the primary and secondary eclipses of close binary 1, and best fit trapezoid model to each of them. The points around 0.968 are the residuals, shifted vertically for visualization. There are significant differences from the simple trapezoid model, but as these are approximately symmetrical from the center of the eclipses, they are not expect to introduce systematic errors in the determination of the time of eclipse centers.} 
\label{fig:fig_pri_sec_model}
\end{figure}

We used a parabolic fit to the baseline in order to remove the modulation between the eclipses, and a Levenberg-Marquardt fit \citep{Press:1992aa} to a trapezoidal function to estimate the parameters of each eclipse (center, depth, duration, and ingress time). Of the four fit parameters for each eclipse, only the central time of the eclipse shows a significant variability, while the other three remain constant. The observed variability has the same period as the wide component $A$. On a final iteration, we corrected for the measured O-C displacement at each time stamp, before computing the average curves for both the primary and secondary eclipse of the inner binary (\emph{templates}), which are plotted in Figure~\ref{fig:fig_pri_sec_model}. For each observed eclipse, the corresponding template was shifted in time on a regular grid, and the \chisq\ between the model and the data was calculated and used to calculate the time of eclipse center and its uncertainty.

\end{appendix}

\end{document}